\documentclass[11pt,twoside]{article}  % Leave intact
\usepackage{adassconf}

\begin{document}   % Leave intact

%-----------------------------------------------------------------------
%			    Paper ID Code
%-----------------------------------------------------------------------
% Enter the proper paper identification code.  The ID code for your
% paper is the session number associated with your presentation as
% published in the official conference proceedings.  You can
% find this number locating your abstract in the printed proceedings
% that you received at the meeting or on-line at the conference web
% site; the ID code is the letter/number sequence proceeding the title 
% of your presentation.  
%
% This will not appear in your paper; however, it allows different
% papers in the proceedings to cross-reference each other.  Note that
% you should only have one \paperID, and it should not include a
% trailing period.
%

\paperID{O12-2}

%-----------------------------------------------------------------------
%		            Paper Title 
%-----------------------------------------------------------------------
% Enter the title of the paper.
%
% EXAMPLE: \title{A Breakthrough in Astronomical Software Development}
%
% If your title is so long as to fill the page header when you print it,
% then please supply a short form as a \titlemark.
%
% EXAMPLE:
%  \title{Rapid Development for Distributed Computing, with Implications
%         for the Virtual Observatory}
%  \titlemark{Rapid Development for Distributed Computing}
%

\title{The AVO to EURO-VO Transition}
%\titlemark{ }

%-----------------------------------------------------------------------
%		          Authors of Paper
%-----------------------------------------------------------------------
% Enter the authors followed by their affiliations.  The \author and
% \affil commands may appear multiple times as necessary.  List each
% author by giving the first name or initials first followed by the
% last name.  Authors with the same affiliations should grouped
% together. 
%
% Try to limit the front matter to no more than three \author
% commands.  Group authors with the same affiliations.  Too many
% \author commands fills the first page of the paper with little
% actual text.

\author{Paolo Padovani, on behalf of the EURO-VO Project}
\affil{European Southern Observatory, Karl-Schwarzschild-Str. 2, 
       D-85748 Garching bei M\"unchen, Germany, 
       Email Paolo.Padovani@eso.org}

%-----------------------------------------------------------------------
%			 Contact Information
%-----------------------------------------------------------------------
% This information will not appear in the paper but will be used by
% the editors in case you need to be contacted concerning your
% submission.  Enter your name as the contact along with your email
% address.

\contact{Paolo Padovani}
\email{Paolo.Padovani@eso.org}

%-----------------------------------------------------------------------
%		      Author Index Specification
%-----------------------------------------------------------------------
% Specify how each author name should appear in the author index.  The 
% \paindex{ } should be used to indicate the primary author, and the
% \aindex for all other co-authors.  You MUST use the following
% syntax: 
%
% SYNTAX:  \aindex{LASTNAME, F. M.}
% 
% where F is the first initial and M is the second initial (if
% used).  This guarantees that authors that appear in multiple papers
% will appear only once in the author index.  
%
% EXAMPLE: \paindex{Crabtree, D.}
%          \aindex{Manset, N.}
%          \aindex{Veillet, C.}
%
% NOTE: this information is also used to build the author list that
% appears in the table of contents.  Authors will be listed in the order
% of the \paindex and \aindex commmands.
%

\paindex{Padovani, P.}

%-----------------------------------------------------------------------
%                     Author list for page header
%-----------------------------------------------------------------------
% Please supply a list of author last names for the page header. in
% one of these formats:
%
% EXAMPLES:
% \authormark{LASTNAME}
% \authormark{LASTNAME1 \& LASTNAME2}
% \authormark{LASTNAME1, LASTNAME2, ... \& LASTNAMEn}
% \authormark{LASTNAME et al.}
%
% Use the "et al." form in the case of seven or more authors, or if
% the preferred form is too long to fit in the header.

\authormark{Padovani}

%-----------------------------------------------------------------------
%			Subject Index keywords
%-----------------------------------------------------------------------
% Enter up to 6 keywords describing your paper.  These will NOT be
% printed as part of your paper; however, they will be used to
% generate the subject index for the proceedings.  There is no
% standard list; however, you can consult the indices for past ADASS
% proceedings (http://adass.org/adass/proceedings/).


%-----------------------------------------------------------------------
%			       Abstract
%-----------------------------------------------------------------------
% Type abstract in the space below.  Consult the User Guide and Latex
% Information file for a list of supported macros (e.g. for typesetting 
% special symbols). Do not leave a blank line between \begin{abstract} 
% and the start of your text.

\begin{abstract}          % Leave intact
The Astrophysical Virtual Observatory (AVO) initiative, jo\-int\-ly funded by
the European Commission and six European organisations, had the task of
creating the foundations of a regional scale infrastructure by conducting a
research and demonstration programme on the VO scientific requirements and
necessary technologies. The AVO project is now formally concluded. I
highlight AVO's main achievements and then describe its successor, the
EURO-VO project. With its three new interlinked structures, the Data Centre
Alliance, the Facility Centre, and the Technology Centre, the EURO-VO is
the logical next step for the deployment of an operational VO in Europe.
\end{abstract}

%-----------------------------------------------------------------------
%			      Main Body
%-----------------------------------------------------------------------
% Place the text for the main body of the paper here.  You should use
% the \section command to label the various sections; use of
% \subsection is optional.  Significant words in section titles should
% be capitalized.  Sections and subsections will be numbered
% automatically. 

\section{Introduction}

The Virtual Observatory (VO) is an innovative, evolving system, which will
allow users to interrogate multiple data centres and services in a seamless
and transparent way, to best utilise astronomical data. The main goal of
the VO is to enable new science by making the huge amount of data presently
on-hand easily accessible to astronomers. Within the VO, data analysis
tools and models, appropriate to deal also with large data volumes, will be
made more available as well. The VO initiative is a global collaboration of
the world's astronomical communities under the auspices of the recently
formed International Virtual Observatory Alliance
(\htmladdnormallinkfoot{IVOA}{http://ivoa.net}).

\section{The Astrophysical Virtual Observatory}

The status of the VO in Europe is very good. In addition to seven
national VO projects members of the IVOA, the European funded
collaborative Astrophysical Virtual Observatory (AVO) project had the
task of creating the foundations of a regional scale infrastructure by
conducting a research and demonstration programme on the VO scientific
requirements and necessary technologies. The AVO was jointly funded by
the Fifth Framework Programme [FP5] of the European Commission 
with six European organisations (ESO, the European Space Agency [ESA],
AstroGrid, the Centre de Donn\'ees astronomique de Strasbourg [CDS],
TERAPIX, and Jodrell Bank) participating in a three year, Phase-A
programme.

The AVO project was driven by a strategy of regular scientific
demonstrations of VO technology and is now formally concluded. AVO's main
achievements can be thus summarised:

\begin{enumerate}

\item {\it Three science demonstrations}. These were held on an annual
basis, in coordination with the IVOA, for the AVO Science Working Group
(SWG), established to provide scientific advice to the project. Three very
successful demonstrations were held in January 2003 (Jodrell Bank), 2004
(ESO, Garching), and 2005 (ESAC, Madrid).

\item {\it First VO paper}. I reported last year (Padovani 2005)
on AVO's second demonstration, held in January 2004 at ESO, which
lead to the discovery of 31 new optically faint, obscured quasar candidates
(the so-called QSO 2) in the two Great Observatories Origins Deep Survey
(GOODS) fields. These results led to the publication of the first
refereed astronomical paper enabled via end-to-end use of VO tools and
systems (Padovani et al. 2004). 

\item {\it Science Reference Mission}. The Science Reference Mission is a
definition of the key scientific results that the full-fledged VO in Europe
should achieve when fully implemented. It consists of a number of science
cases, with related requirements, against which the success of the
operational VO in Europe will be measured. It was put together by the AVO
SWG.

\item {\it VO tools}. Progressively more complex AVO demonstrators
have been constructed. The current one is an evolution of Aladin, developed
at CDS, and has become a set of various software components, provided
by AVO and international partners, which allows relatively easy access
to remote data sets (images and spectra), manipulation of image and
catalogue data, and remote calculations in a fashion similar to remote
computing. The AVO prototype is a VO tool which can be used now for
the day-to-day work of astronomers and can be downloaded from the AVO
Web site as a Java application. We note that this is by definition a
prototype and will not be maintained on the long term. Most of the
functionalities developed for the AVO demonstrations are now available
in the public version of Aladin, and the inclusion of the remaining
ones is being assessed.

\item {\it VO development.} AVO provided fundamental input to the IVOA
for the development and the usage of the following VO standards: VOTable,
Data Access Layer, Data Model, Uniform Content Descriptor, and Web
Services.

\item {\it Creation of IVOA.} AVO has been one of the founding member of
the IVOA.

\end{enumerate}

Links to various documents and
to the software download page can be found at
\htmladdURL{http://www.euro-vo.org/twiki/bin/view/Avo/}.

\section{The EURO-VO}

The EURO-VO work program is the logical next step from AVO as a Phase-B
deployment of an operational VO in Europe (see Hanisch 2006 for similar
transitional activities in other VO projects). Building on the development
experience gained within the AVO Project, in coordination with the European
astronomical infrastructural networks OPTICON and RADIONET, and through
membership and support of the IVOA, EURO-VO will seek to obtain the
following objectives:

\begin{enumerate}

\item technology take-up and full VO compliant data and
resource provision by astronomical data centres in Europe; 

\item support to the scientific community to utilise the new VO
infrastructure through dissemination, workshops, project support, and VO
facility-wide resources and services;

\item building of an operational VO infrastructure in response
to new scientific challenges via development and refinement of VO
components, assessment of new technologies, design of new components and
their implementation. 
\end{enumerate}

EURO-VO is open to all European astronomical data centres. Initial partners
include ESO, the European Space Agency, and six national funding agencies,
with their respective VO nodes: Istituto Nazionale di Astrofisica (INAF,
Italy), Institut National des Sciences de l'Univers (INSU, France),
Instituto Nacional de Tecnica Aeroespacial (INTA, Spain), Nederlandse
Onderzoekschool voor Astronomie (NOVA, Netherlands), Particle Physics and
Astronomy Research Council (PPARC, UK), and Rates Deutscher Sternwarten
(RDS, Germany). The total planned EURO-VO resources sum up to approximately
60 persons/yr over three years, i.e., about three times those of the AVO.

EURO-VO will seek to obtain its objectives by establishing three new
interlinked structures: 

\begin{enumerate}

\item the EURO-VO Data Centre Alliance (DCA), an alliance of European data
centres who will populate the EURO-VO with data, provide the physical
storage and computational fabric and who will publish data, metadata and
services to the EURO-VO using VO technologies;

\item the EURO-VO Facility Centre (VOFC), an organisation that provides the
EURO-VO with a centralised registry for resources, standards and
certification mechanisms as well as community support for VO technology
take-up and dissemination and scientific program support using VO
technologies and resources;

\item the EURO-VO Technology Centre (VOTC), a distributed organisation that
coordinates a set of research and development projects on the advancement
of VO technology, systems and tools in response to scientific and community
requirements.

\end{enumerate} 

The DCA will be a persistent alliance of data centre communities
represented at a national level. Through membership in the DCA, a nation's
community of data curators and service providers will be represented
in a forum that will facilitate the take-up of VO standards, share best
practice for data providers, consolidate operational requirements for
VO-enabled tools and systems and enable the identification of
scientific requirements from programs of strategic national interest that
require VO technologies and services. Funds for the DCA have been requested
in an FP6 proposal submitted in September 2005.

The VOFC will provide a "public face" to the EURO-VO. Through outreach,
support of VO-enabled science projects in the community, workshops and
schools, the VOFC will represent a central support structure to facilitate
the broad take-up of VO tools by the community. The VOFC will also support
the EURO-VO Science Advisory Committee (SAC) to ensure appropriate and
effective scientific guidance from the community of leading researchers
outside the mainstream VO projects. The SAC will provide an up-to-date
stream of high-level science requirements to the EURO-VO. The VOFC will
further provide central services to the DCA for resource registry, metadata
standards and EURO-VO access. Funding for the VOFC has yet to be fully
defined but will come initially from ESO and ESA with activities ramping up
in 2006. The first VOFC activity was the organisation of the
\htmladdnormallinkfoot{EURO-VO
workshop}{http://wwww.euro-vo.org/workshop2005} at ESO Headquarters in
Garching from June 27 to July 1, 2005. 

The VOTC will consist of a series of coordinated technology research and
development projects conducted in a distributed manner across the member
organisations. The first project under the VOTC is the
\htmladdnormallinkfoot{VO-TECH project}{http://eurovotech.org} (Walton et
al. 2006), funded through the EC FP6 Proposal and contributions from the
Universities of Edinburgh, Leicester, and Cambridge in the United Kingdom,
ESO, CNRS and Universit\'e Louis Pasteur (France), and INAF (Italy).
Additional projects can be brought to the VOTC via other member
organisations; one such example is ESA-VO. The VOTC provides a mechanism
to coordinate and share technological developments, a channel for DCA and
VOFC requirements to be addressed and for technological developments to be
distributed to the community of data centres and individual scientists in a
coordinated and effective manner.

The EURO-VO project will be proactive in reaching out to European
astronomers. The EURO-VO has also started making regular appearances
at Joint European and National Astronomy Meetings (JENAM), as of the
one in Liege in July 2005. Moreover, the EURO-VO will also help data
centres beyond the partners' countries to join the VO effort.

As an example of partners' involvement, I note that on November 1, 2004,
the Data Management and Operations Division at ESO has created the Virtual
Observatory Systems (VOS) Department. VOS' role is to manage ESO's VO
activities and to make its Science Archive Facility into a powerful
scientific resource for the ESO community by creating, ingesting, and
publishing advanced data products, that is high-level (or "science-grade")
data (Rosati 2006).

% You can also add an acknowledgments section as indicated below.

\acknowledgments 
The Astrophysical Virtual Observatory was selected for funding by the
Fifth Framework Programme of the European Community for research,
technological development and demonstration activities, under contract
HPRI-CT-2001-50030. The EURO-VO VO-TECH project was selected for
funding by the Sixth Framework Programme of the European Community.

\end{document}